\documentclass[aps,pra,superscriptaddress,amsmath,amssymb,preprintnumbers,showpacs]{revtex4}

\usepackage{graphicx}% Include figure files
\usepackage{dcolumn}% Align table columns on decimal point
\usepackage{bm}% bold math
\usepackage{amsmath}
\begin{document}
\title{Quantum superpositions of clockwise and counterclockwise supercurrent states in the  dynamics of a rf-SQUID
exposed to a quantized electromagnetic field}
\author{R. Migliore}
\email{rosanna@fisica.unipa.it}

\affiliation{INFM and MIUR, Dipartimento di Scienze Fisiche ed
Astronomiche dell'Universit\`{a} di Palermo, via Archirafi 36,
90123 Palermo, Italy.}

\author{A. Messina}

\affiliation{INFM and MIUR, Dipartimento di Scienze Fisiche ed
Astronomiche dell'Universit\`{a} di Palermo, via Archirafi 36,
90123 Palermo, Italy.}

\date{\today}

\begin{abstract}
The dynamical behavior of a superconducting quantum interference
device (a rf-SQUID) irradiated by a single mode quantized
electromagnetic field is theoretically investigated. Treating the
SQUID as a flux qubit, we analyze the dynamics of the combined
system within the low lying energy Hilbert subspace both in the
asymmetric and in the symmetric SQUID potential configurations. We
show that the temporal evolution of the system is dominated by an
oscillatory behavior characterized by more than one, generally
speaking, incommensurable Rabi frequencies whose expressions are
explicitly given. We find that the external parameters may fixed
in such a way to realize a control on the dynamical replay of the
total system which, for instance, may be forced to exhibit a
periodic evolution accompanied by the occurrence of  an
oscillatory disappearance  of entanglement between the two
subsystems. We demonstrate the possibility of generating quantum
maximally entangled superpositions of the two macroscopically
distinguishable states describing clockwise and counterclockwise
supercurrents in the loop. The experimental feasibility of our
proposal is briefly discussed.
\end{abstract}


\maketitle

\section{Introduction}
It is of great interest to understand and study the physics of
Josephson junction-based devices both for testing fundamental
properties of quantum mechanics, such as the superposition
principle or the occurrence of entangled states \cite{legg}, and
for technological applications in the context of quantum
information theory and quantum computing \cite{divi}.

In the last decade, rapid developments in the realm of
nanotechnologies have made it possible to perform a number of
beautiful and sophisticated experiments at low temperature
\cite{rouse} bringing to light the existence in these
\emph{atom-like} circuits of many macroscopic quantum phenomena
like energy level quantization \cite{clarke}, macroscopic quantum
tunneling (MQT) and quantum superposition of states
\cite{luk2000,van der}. More recently, the usefulness of
investigating these solid state devices in the context of quantum
communication and information theory has been fully recognized.

Superconducting Josephson devices may, in fact, be thought of as
two-state systems realizing the elementary unit of quantum
information, known as quantum bit or qubit \cite{weiss}. Moreover,
Josephson devices can be scaled up to a large number of qubits and
their dynamics may be controlled by externally applied voltages
and magnetic fluxes. Superconducting devices like Cooper pair
boxes, Josephson junctions or SQUIDs have been thus proposed and
used as basic elements for the practical realization of quantum
gates and chips \cite{makhlin}. The relevant macroscopic degree of
freedom, allowing to store and manipulate quantum information, may
be the charge on the island of a Cooper pair box or the phase
differences at the junction. In the first case the charging energy
$E_C$ overcomes the Josephson energy $E_J$. Otherwise, in the
opposite regime, the Josephson energy overcomes $E_C$ \cite{libro
lik}. It has been already experimentally demonstrated that Cooper
pair boxes behave as two level systems which can be coherently
controlled \cite{naka,nakamu,devoret} and now great efforts are
devoted to prove that the same can be done with flux and phase
qubits \cite{luk2000,van der, mart}. However, successful
realization of quantum algorithms critically depends on the
ability to entangle quantum states of qubits. The optimum would be
the realization of a tunable coupling bus. Several coupling
mechanisms are possible but the natural way of coupling two or
more superconducting qubits is through an intermediate
\emph{resonant LC circuit}, playing the role of a data bus
\cite{plastica,hekk}. Such a resonant LC circuit, describable as a
quantum harmonic oscillator, may be in principle replaced by the
electromagnetic single-mode of a high-Q cavity or by a large-area
current biased Josephson junction. In all these cases the
situation is similar to cavity QED \cite{cqed} (where cavity and
atoms play the roles of the LC circuit and qubits, respectively)
and to ion-trap proposals \cite{iontrap}.

It is then evidently of interest to study the interaction between
a two level solid state system, like an rf-SQUID, and an external
quantum system like another qubit, a tank circuit or a
monochromatic radiation source
\cite{EPJB,EPJbb,stroud,buisso,hekk,digg,ever,vietri}. The aim of
such investigations is to bring to light the occurrence of
entangled states and to construct coupling schemes by which the
coherent dynamics of the system may be controlled and/or
manipulatedì \cite{Yu,maki,Han2,plastica,mart,naka,devoret}.

In this paper our main scope is to study the dynamics of a flux
qubit (an rf-SQUID) coupled to a single mode quantized
electromagnetic field of a resonant cavity. Confining ourselves to
the low-lying energy Hilbert space, we prove that the time
evolution of the combined system is characterized by the
occurrence of entanglement which may be controlled in terms of the
strength of the coupling and the circuit parameters. In addition
we show that the dynamics is dominated by an oscillatory behavior
traceable back to the existence of a finite set of characteristic
Rabi frequencies whose expression may be explicitly given.

The importance of conceiving experimental schemes for realizing
quantum superpositions of macroscopically distinguishable states
has been quite recently emphasized \cite{legg2002}. The main
result of this paper is that, appropriately acting upon some
control parameters, it is possible to guide the rf-SQUID toward
coherent maximally entangled combinations of two states describing
clockwise and counter clockwise supercurrents and then
macroscopically distinguishable.

In section II we describe the physical system under study. Its
dynamics in a reduced low-lying energy Hilbert space is studied in
section III where the main results of this paper are reported. In
the last section, we discuss our results and we conclude with some
remarks about the feasibility of an experiment aimed at verifying
our theory in the laboratory.

\section{The quantum circuit}
In this section we describe in detail the physical system, namely
a rf-SQUID coupled to a monochromatic field of a high-Q resonant
cavity and its hamiltonian model. In figure 1 the electromagnetic
single-mode cavity is represented as an LC resonator.

In subsection II A we describe the physical conditions that allow
us to consider a SQUID as a two level system, that is as a flux
qubit. Then we give the hamiltonian model for the cavity field in
terms of the combined system characteristics and finally, in
subsection II C, we consider the inductive coupling between these
two subsystems.

\begin{figure}
\begin{center}
\includegraphics{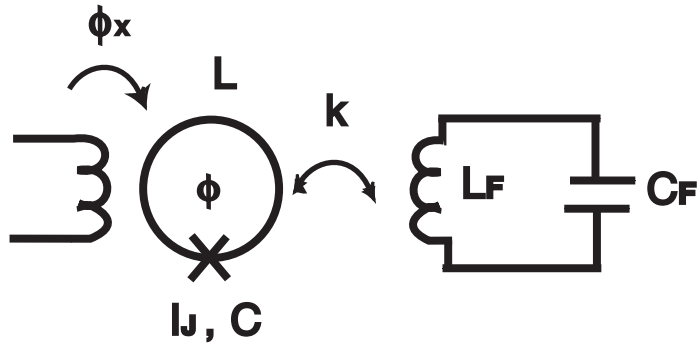}
\caption{\footnotesize Schematics of the superconducting circuits
with a flux qubit inductively coupled to a LC resonator modelling
a single-mode quantized electromagnetic field of a resonant high-Q
cavity.}
\end{center}
\end{figure}

\subsection{ The rf-SQUID as material two level system}
Let us begin by describing as usual \cite{libro lik} the rf-SQUID,
a superconducting loop interrupted by a Josephson junction (figure
2a), as a fictitious particle of mass $C$ and generalized
coordinate $ \phi$ (the magnetic flux in the loop) subjected to
the washboard potential
\begin{eqnarray} \label{pote}
U(\phi)=-E_J \;\cos \left(2 \pi \frac{\phi}{\phi_0}\right)+
\frac{(\phi- \phi_x)^2}{2L}
\end{eqnarray}
where $C \sim 10^{-15} \div 10^{-13}\;F$ is the junction
capacitance, $L \sim 10 \div 100\;pH$ the self-inductance of the
loop, $\phi_x$ an externally applied dc flux and $\phi_0=
\frac{h}{2e}$ the flux quantum. The Josephson coupling energy
$E_J$ is related to the critical supercurrent $I_C$ by $E_J=I_C
\phi_0/2 \pi$.

Taking into account both the kinetic and potential energy, it is
then immediate to write the Hamiltonian of the system as follows
\begin{eqnarray}\label{hsquid}
H=\frac{Q^2}{2C}-E_J \;\cos \left(2 \pi
\frac{\phi}{\phi_0}\right)+ \frac{(\phi- \phi_x)^2}{2L}
\end{eqnarray}
where the charge on the junction capacitance $Q=-i \hbar
\partial/ \partial \phi$ and the flux $\phi$ in the loop are canonically
conjugate operators satisfying the commutation rule $\lbrack \phi,
Q\rbrack=i \hbar$. Under the condition $\beta_L=2 \pi L I_C/\phi_0
> 1$ the rf-SQUID is hysteretic and the potential $U(\phi)$ may
have one or several relative minima. The height of barrier between
minima and the number of minima depend on the parameter $\beta_L$.
The form of the potential $U(\phi)$ instead can be tuned by
changing the external dc magnetic flux $\phi_x$ applied to the
loop to the case where the two lowest energy wells are degenerate.
This case occurs for $\phi_x=\phi_0/2$. These two degenerate
minima correspond to the clockwise and counterclockwise sense of
rotation of the supercurrent in the loop. The two relative ground
states are localized flux states, hereafter denoted $\vert L
\rangle$ and $\vert R \rangle$. For $\phi_x$ not exactly
coincident with $ \phi_0/2$, $U(\phi)$ defines an asymmetric
double-well potential near $\phi=\phi_0/2$ with barrier $V_b$
between these two lowest energy minima. At sufficiently low
temperatures, transitions between neighboring flux states are
dominated by macroscopic quantum tunnelling. However, for high
barrier $V_b$, tunnelling does not mix the two lowest flux states
with the excited states in the two wells. Thus, in the parameter
regime $V_b \gg \hbar \omega_0 \gg k_BT$ ( $\hbar \omega_0$ being
the separation of the first excited state from the ground state in
both wells), the rf-SQUID effectively behaves as a two state
system \cite{weiss} with reduced Hamiltonian expressible in terms
of the Pauli matrices $\sigma_x$ and $\sigma_z$ as follows
\begin{eqnarray}\label{hsqui2}
H_{S}= -\frac{\hbar}{2} \;\Delta \sigma_x+\frac{\hbar}{2} \;
\epsilon \sigma_z= \frac{\hbar}{2} \left(
\begin{array}{c c}
\epsilon &-\Delta \\
-\Delta &- \epsilon \\
\end{array}
\right),
\end{eqnarray}
where the basis coincides with the two localized flux states
$\vert L \rangle$ and $\vert R \rangle$. Here $\hbar
\epsilon(\phi_x) \equiv \hbar \epsilon=2I_C
\sqrt{6(\beta_L-1)}(\phi_x- \phi_0/2)$ is the asymmetry of the
double-well and $\Delta$ is the tunnelling frequency between the
wells. This tunnelling frequency can be tuned by changing the
height $V_b$ of  the barrier which depends on the Josephson
coupling energy $E_J$. Then if we replace the junction by a
hysteretic dc-SQUID (Fig. 2b), behaving as a JJ with tunable
critical current, $E_J$ may be manipulated by a separately control
dc flux $\phi_c$. In other words, this modified device, known as
the double rf-SQUID, behaves as a normal rf-SQUID whose Josephson
energy $E_J$ is related to the control flux $\phi_c$ by
$E_J(\phi_c)=\frac{\phi_0}{\pi}I_{C0} \cos (\pi
\frac{\phi_c}{\phi_0}$), $I_{C0}$ being the critical current of
each of the two JJs of the dc-SQUID.

\begin{figure}
\begin{center}
\includegraphics{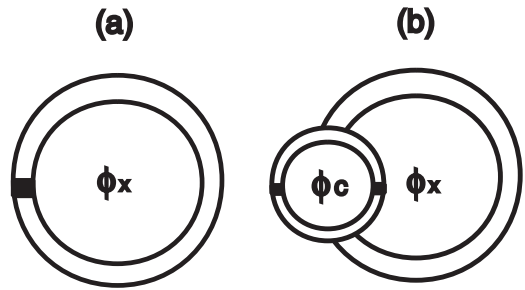}
\caption{\footnotesize \textbf{(a)} Schematics of a rf-SQUID  and
\textbf{(b)} of a double rf-SQUID  obtained replacing the single
Josephson junction with a dc-SQUID, namely a superconducting loop
interrupted by two Josephson junction. }
\end{center}
\end{figure}
Since in our scheme we need to control $\Delta$, in what follows,
we will investigate a double rf-SQUID with two external control
fluxes $\phi_x$ and $\phi_c$ when it is exposed to a monochromatic
quantized electromagnetic field.

The hamiltonian of the double rf-SQUID, given also in this case by
eq. \eqref{hsqui2}, can be easily cast in diagonal form
\begin{eqnarray}\label{hsdiag}
H_S= \frac{\hbar}{2} \left(
\begin{array}{c c}
-\sqrt{\epsilon^2+\Delta^2} &0 \\
0&\sqrt{\epsilon^2+\Delta^2} \\
\end{array}
\right)
\end{eqnarray}
in the basis formed by its eigenstates
\begin{eqnarray}\label{auto-}
\vert- \rangle=C_- \lbrack \vert R
\rangle+\frac{\epsilon+\sqrt{\epsilon^2+\Delta^2}}{\Delta}\vert L
\rangle \rbrack
\end{eqnarray}
and
\begin{eqnarray}\label{auto+}
\vert+ \rangle=C_+ \lbrack \vert R
\rangle+\frac{\epsilon-\sqrt{\epsilon^2+\Delta^2}}{\Delta}\vert L
\rangle \rbrack.
\end{eqnarray}
Here $C_{\mp}=\lbrack 1+\left(\frac{\epsilon \pm
\sqrt{\epsilon^2+\Delta^2}}{\Delta}\right)^2\rbrack^{-\frac{1}{2}}$
are the normalizing factors of $\vert-\rangle$ and $\vert
+\rangle$ respectively and $\hbar \sqrt{\epsilon^2+\Delta^2}$ is
the energy difference between their corresponding eigenvalues
$E_-=-\frac{\hbar}{2} \sqrt{\epsilon^2+\Delta^2}$ and
$E_+=\frac{\hbar}{2} \sqrt{\epsilon^2+\Delta^2}$.

\subsection{ The quantized electromagnetic field}

In this section we model the electromagnetic field of the resonant
cavity as a $LC$ resonator. In this way our treatment and our
results may be easily extended to the case of the coupling between
the flux qubit and a real $LC$ resonator \cite{buisso} or a
large-area current-biased Josephson junction \cite{hekk,mart}.
This fact is important because, in order to verify experimentally
our theoretical predictions, it is more practical to couple the
qubit with one of these two systems rather than with the
single-mode of a resonant cavity. We therefore start by
considering the Hamiltonian of an $L_FC_F$ resonator with infinite
parallel resistance on resonance and frequency $\omega_F=1/
\sqrt{L_FC_F}$
\begin{eqnarray}\label{hreso}
H_F=\frac{Q_F^2}{2C_F}+\frac{\phi_F^2}{2L_F}.
\end{eqnarray}
where $\phi_F$ and $Q_F$ play the roles of the magnetic flux and
charge operators arising from the quantized electromagnetic field
and satisfying the commutation rule $\lbrack \phi_F,Q_F\rbrack=i
\hbar$.

As in references \cite{EPJB,stroud} it is possible to relate the
resonator operators $\phi_F$ and $Q_F$ to bosonic annihilation and
creation operators $a$ and $a^{\dagger}$. Defining the flux
$\phi_F$ and the conjugate operator $Q_F $ as

\begin{eqnarray}\label{fluxfield RES}
\phi_F=\sqrt{ \frac{\hbar}{2 \omega_FC_F}}\;(a+a^{\dagger})
\end{eqnarray}
and

\begin{eqnarray}\label{Qfield}
Q_F=-i \sqrt{ \frac{\hbar \omega_F C_F}{2 }}\;(a-a^{\dagger})
\end{eqnarray}
and substituting these analytical expressions in eq.
\eqref{hreso}, it is immediate to cast the Hamiltonian of the
resonator in the form of a standard harmonic oscillator free
Hamiltonian

\begin{eqnarray}\label{hoscarm}
H_F=\hbar \omega_F(a^{\dagger}a+\frac{1}{2}).
\end{eqnarray}
Here the frequency $\omega_F$, choosing $C_F\sim 10^{-12}\; F$ and
$L_F \sim 0.1 \;nH$, belongs in the range of microwaves ($\omega_F
\approx 10^{11} \;rad \cdot s^{-1}$). The eigenfunctions of
hamiltonian \eqref{hoscarm} are, of course, harmonic oscillator
eigenstates $\vert n \rangle$ (defined by $a^{\dagger}a \vert n
\rangle=n \vert n \rangle$) with eigenvalues $E_n=\hbar
\omega_F(n+\frac{1}{2})$.

If the rf-SQUID is effectively coupled with the quantized mode of
an electromagnetic high-Q cavity, we may start again from
Hamiltonian \eqref{hreso}. In order to understand the meaning of
$C_F$ in this case, we assume that the rf-SQUID is located
perpendicularly to the magnetic field and within a distance small
compared to the radiation wavelength. In such conditions, the
vector potential $\mathbf{A}(\mathbf{x})$ arising from the
electromagnetic field of the cavity mode is approximately uniform
throughout the region of the device and in international units and
in the Coulomb gauge ($\nabla \cdot\mathbf{A}=0$) it assumes the
form:
\begin{eqnarray}\label{poteA}
\mathbf{A}=\left(\frac{\hbar}{2 \epsilon_0
\omega_FV}\right)^{\frac{1}{2}}(a-a^{\dagger})\;\mathbf{u},
\end{eqnarray}
$\mathbf{u}$ being the unit polarization vector, $\epsilon_0$ the
vacuum dielectric constant and $V$ the quantization volume of the
field mode. Thus the expression for the operator $\phi_F$ to be
inserted in eq. \eqref{hreso} may be written down as
\begin{eqnarray}\label{fluxfield}
\phi_F=\oint_{\gamma}\mathbf{A} \cdot d\mathbf{l}=\sqrt{
\frac{\hbar}{2 \omega_FC_F}}\;(a+a^{\dagger})
\end{eqnarray}
where the capacitive parameter $C_F$, given by the following
expression
\begin{eqnarray}\label{cf}
C_F=\epsilon_0 V\left(\oint_{\gamma}\mathbf{u}\cdot d
\mathbf{l}\right)^{-2},
\end{eqnarray}
depends on the field frequency, via the quantization volume $V$,
and on the SQUID geometry, via the line integral which is taken
across a closed circuit $\gamma$ inside the SQUID loop. Then, also
in this case, exploiting eq. \eqref{fluxfield} and the properties
of conjugation between $\phi_F$ and $Q_F$, it is easy to cast
Hamiltonian \eqref{hreso} in the form of a standard harmonic
oscillator free Hamiltonian \eqref{hoscarm}.

\subsection{The coupled system}
In view of the assumptions made in the previous subsection, the
flux qubit and the $LC$-resonator modelling the monochromatic
field can be thought of as coupled together inductively (see
figure 1) with a contribution to the total hamiltonian given by
\begin{eqnarray}\label{hint}
H_I=\frac{2k}{L}\phi \phi_F=B(a+a^{\dagger})[-\epsilon \sigma_z+
\Delta \sigma_x ]
\end{eqnarray}
where the constant
\begin{eqnarray}\label{B}
B=\frac{k}{L}\sqrt{\frac{\hbar}{2 \omega_FC_F}}
\frac{\phi_0}{\sqrt{\epsilon^2+\Delta^2}}
\end{eqnarray}
has the same dimension of $\hbar$ and depends on the coupling
strength  and the system characteristics. Here the flux linkage
factor $k$ is assumed of the order of $0.01$ in accordance with
the current experimental values \cite{chiarello}. Thus the
Hamiltonian describing the combined rf-SQUID-field system can be
written down as
\begin{eqnarray}\label{htot}
H=H_S+H_F+H_I.
\end{eqnarray}
where $H_S$ and $H_F$, given by eq. \eqref{hsdiag} and
\eqref{hoscarm} are the Hamiltonians describing the two free
subsystems.

\section{The system dynamics in a reduced low lying energy $4 \times 4$ Hilbert
subspace}

In view of eqs. \eqref{hsdiag}, \eqref{hoscarm} and \eqref{hint}
the Hamiltonian $H$ may be formally interpreted as that of a
two-level \lq \lq atom\rq \rq \;interacting quantum mechanically
with a monochromatic quantized electromagnetic field. We wish now
to focuss our attention on a physical situation wherein the
operating temperature is $\;T \approx 10 \;mK$ and the field is in
resonance with the transition  between the lowest and the first
excited state of the rf-SQUID, that is $\hbar
\omega_F=\hbar\sqrt{\epsilon^2+\Delta^2}$.

As a consequence we drastically simplify the difficult problem of
diagonalizing $H$ in its infinite dimensional Hilbert space by a
truncation procedure which confine ourselves to the $4 \times 4$
Hilbert space spanned by the states $\lbrace \vert 0-
\rangle,\vert 1- \rangle ,\vert 0+ \rangle, \vert 1+
\rangle\rbrace $. In other words, we investigate the dynamics of
the system in the low-lying energy subspace of the free
Hamiltonian $H_0=H_S+H_F$ generated by the four states, $\vert n
\sigma \rangle$ ($n \equiv 0,1; \; \sigma \equiv -,+ $) which are
the common eigenstates of $H_S$ and $H_F$ such that $H_S \vert n
\sigma \rangle= E_{\sigma} \vert n \sigma \rangle $ and $H_F \vert
n \sigma \rangle=\hbar \omega_F(n+ \frac{1}{2}) \vert n \sigma
\rangle$.

In this reduced basis and adopting the resonant condition
$\omega_F=\sqrt{\epsilon^2+\Delta^2}$, Hamiltonian \eqref{htot}
can be cast in the form of the following $4 \times 4$ non diagonal
matrix:
\begin{eqnarray}\label{hasimm}
H_R= \left( \begin{array}{c c c c}

0 & -B \epsilon & 0 & B \Delta  \\

-B \epsilon &  \hbar \omega_F & B \Delta & 0\\

0 & B \Delta & \hbar \omega_F & B \epsilon \\

B \Delta & 0 &  B \epsilon &2 \hbar \omega_F

\end{array}
\right).
\end{eqnarray}
We emphasize that the counter-rotating terms of $H$ contribute to
this truncated Hilbert space with the presence of non diagonal
matrix elements of $H_R$. These matrix elements are those
connecting states of the basis having a different number $n+
\sigma$ of total excitations.

In the following, we wish to study the dynamics of the coupled
matter-radiation system initially prepared in different physically
meaningful conditions both in the asymmetric ($\epsilon\neq 0$)
and symmetric ($\epsilon=0$) cases. To this end we need to find
the eigensolutions of $H_R$.

\subsection{\label{sec:level2} The dynamics of the system in the asymmetric case}
Let us consider the time evolution of the combined
rf-SQUID-radiation system initially prepared in the state $\vert0+
\rangle$, that is the field in its vacuum state $\vert0 \rangle$
and the SQUID in the first excited state $\vert+ \rangle =C_+
\lbrack \vert R
\rangle+\frac{\epsilon-\sqrt{\epsilon^2+\Delta^2}}{\Delta}\vert L
\rangle \rbrack$ which is a superposition of the localized flux
states $\vert L \rangle$ and $\vert R \rangle$.

It is worth noting that the manipulation of the shape of the
potential and the height of the barrier between the two lowest
energy wells via the control fluxes $\phi_x$ and $\phi_c$ allows
to prepare the SQUID in a prefixed initial condition
\cite{makhlin,chiarello}. At the same time, exploiting the
currently available experimental techniques in CQED, it is
possible to prepare the field mode in a Fock state with $0$ or $1$
photon \cite{raimond}. Eigenvalues $\vert u_i\rangle$ ($i \equiv
1,2,3,4$) and eigenvectors $\lambda_i$ of $H_R$ in the asymmetric
case may be exactly evaluated and they are explicitly given in
Appendix A.

The expansion of $\vert0 +
\rangle_t=\exp(-i\frac{H_Rt}{\hbar})\;\vert0 + \rangle$ in terms
of $\vert u_1 \rangle$, $\vert u_2 \rangle$, $\vert u_3 \rangle$
and $\vert u_4 \rangle$ may be cast in the following form:
\begin{multline}\label{0+asimmt}
\vert0+ \rangle_ = \;\frac{B \epsilon}{(P_1-P_2)Q_1Q_2}\;\lbrack
\sqrt{n_1}(P_2-Q_1)Q_2 \;\vert u_1 \rangle \exp(-i \lambda_1 t)
-\sqrt{n_2}(P_2+Q_1)Q_2\;\vert u_2 \rangle \exp(-i \lambda_2 t)+
\\-\sqrt{n_3}(P_1-Q_2)Q_1 \;\vert u_3 \rangle \exp(-i \lambda_3 t)+
\sqrt{n_4}(P_1+Q_2)Q_1 \;\vert u_4\rangle \exp(-i \lambda_4
t)\rbrack.
\end{multline}
Thus $\vert 0+ \rangle$ is as in linear superposition, with
different weights, of the eigenstates $\vert u_i \rangle$ of
Hamiltonian \eqref{hasimm}.

We are interested in exploring the ability of the system to
periodically come back to the initial state $\vert 0+ \rangle$ as
well as to pass thought the other states $\vert 1-\rangle$, $\vert
0-\rangle$ and $\vert 1+ \rangle$. The structure of the previous
equation and those of eqs. \eqref{u1}-\eqref{u4} makes clear that
the time evolution of the system from $\vert 0+ \rangle$ involves
all the four states $ \vert 0- \rangle$, $\vert 1- \rangle$,
$\vert 0+ \rangle$ and $\vert 1+ \rangle$. However, since the
states $\vert 0+ \rangle$ and $\vert 1- \rangle$ are almost
degenerate in energy, in our physical situation transitions
between them are more probable than transitions between the
initial state $\vert0+ \rangle$ and the states $\vert 0- \rangle$
and $\vert 1+ \rangle$. This fact may be clearly appreciated by
evaluating the time evolution of the survival probability $P_1(t)$
of the state $\vert 0+ \rangle$, that is
\begin{multline}\label{P0+asimm}
P_1(t)=\vert \langle0+\vert0+ \rangle_t \vert^2=S^2 \lbrace
\sum^4_{j=1}S_j^2+2S_1S_2 \; \cos\frac{Q_1}{\hbar}t+2S_3S_4 \;
\cos \frac{Q_2}{\hbar}t+\\+2 \lbrack S_1S_3+S_2S_4\rbrack \;\cos
\frac{(Q_1-Q_2)}{2 \hbar}t+ 2 \lbrack S_2S_3+S_1S_4\rbrack \; \cos
\frac{(Q_1+Q_2)}{2\hbar}t\rbrace
\end{multline}
where
\begin{eqnarray}
S=\frac{B \epsilon}{(P_1-P_2)Q_1Q_2}
\end{eqnarray}
\begin{eqnarray}
S_1=\frac{P_1Q_2}{2 B \epsilon}(Q_1-P_2)
\end{eqnarray}
\begin{eqnarray}
S_2=\frac{P_1Q_2}{2 B \epsilon}(Q_1+P_2)
\end{eqnarray}
\begin{eqnarray}
S_3=-\frac{P_2Q_1}{2 B \epsilon}(Q_2-P_1)
\end{eqnarray}
and
\begin{eqnarray}
S_4=-\frac{P_2Q_1}{2 B \epsilon}(Q_2+P_1).
\end{eqnarray}

It is immediate to construct explicit expressions also for the
transition probabilities $P_2(t)=\vert \langle1-\vert0+ \rangle_t
\vert^2$, $P_3(t)=\vert \langle0-\vert0+ \rangle_t \vert^2$ and
$P_4(t)=\vert \langle1+\vert0+ \rangle_t \vert^2$ to the states
$\vert 1-\rangle$, $\vert 0- \rangle$ and $\vert 1+ \rangle$
respectively. Their explicit analytical expressions are given in
appendix B.  Here we wish to underline that these transition
probabilities have the same mathematical structure of eq.
\eqref{P0+asimm}. This means that $P_2$, $P_3$ and $P_4$ like
$P_1$ are given by the sum of a constant and $4$ trigonometric
time dependent terms with different weights and frequencies. Since
these four frequencies $\omega_1\equiv\frac{Q_1}{\hbar}$,
$\omega_2\equiv \frac{Q_2}{\hbar}$, $\omega_3 \equiv
\frac{Q_1-Q_2}{2 \hbar}$ and $\omega_4\equiv
\frac{Q_1+Q_2}{2\hbar}$ appearing in the expressions of $P_1(t)$,
$P_2(t)$, $P_3(t)$ and $P_4(t)$ are in general incommensurable, we
find a quasi-periodic behavior wherein a complete exact inversion
of the populations between the degenerate states $\vert0+\rangle$
and $\vert 1- \rangle$ never occurs. The time evolutions of $P_1$,
$P_2$, $P_3$ and $P_4$ are plotted in figure 3.

\begin{figure}
\includegraphics{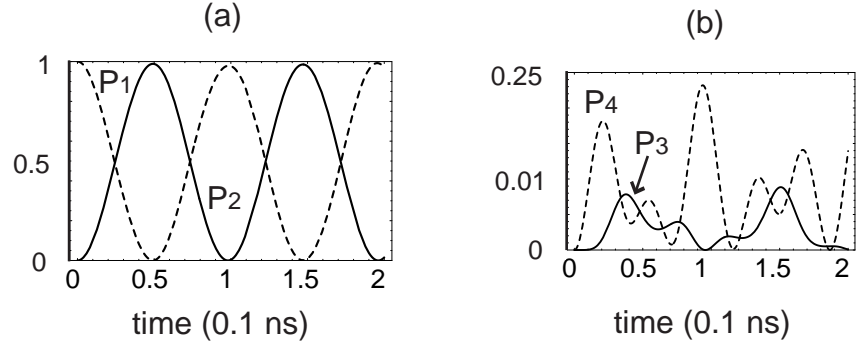}
\caption{\footnotesize \textbf{(a)} Survival probability $P_1$ of
the state $\vert 0+ \rangle$ (dashed line) and transition
probabilities $P_2$, \textbf{(b)} $P_3$ and $P_4$ (dashed line) to
the states $\vert 1- \rangle$, $\vert 0- \rangle$ and $\vert 1+
\rangle$ respectively, for a system with $\epsilon \approx 3 \cdot
10^{10} \;rad \cdot s^{-1}, \;B \approx 2,7 \cdot 10^{-35} J \cdot
s $ and $\omega_F \approx 10^{11} rad \cdot s^{-1}  $.}
\end{figure}

A similar quasi-periodic behavior characterizes the time evolution
of the system initially prepared in the state $\vert 0R
\rangle=\vert 0 \rangle \otimes \vert R \rangle$, that is the
field in the vacuum state $\vert0 \rangle$ and the rf-SQUID in the
localized flux state $\vert R \rangle$ characterized by a well
defined sense of circulation of the supercurrent in the loop. In
view of the definitions of $\vert - \rangle$ and $\vert + \rangle$
as well as of eqs. \eqref{0-a} and \eqref{0+a}, the survival
probability of the state $\vert 0R \rangle$ can be cast in the
following form
\begin{multline}\label{P0Rasimm}
P_{0R}(t)=\vert \langle0R\vert0R \rangle_t \vert^2=W^2 \lbrace
\sum^4_{j=1}W_j^2+2W_1W_2 \; \cos\frac{Q_1}{\hbar}t+2W_3W_4 \;
\cos \frac{Q_2}{\hbar}t+\\+2 \lbrack W_1W_3+W_2W_4\rbrack \;\cos
\frac{(Q_1-Q_2)}{2 \hbar}t+ 2 \lbrack W_2W_3+W_1W_4\rbrack \; \cos
\frac{(Q_1+Q_2)}{2\hbar}t\rbrace
\end{multline}
where now
\begin{eqnarray}
W=\frac{B }{(P_1-P_2)Q_1Q_2}
\end{eqnarray}
\begin{eqnarray}
W_1=Q_2 \lbrack \epsilon \delta_+(P_2-Q_1)+P_2 \Delta \delta_-
\rbrack(-\frac{P_2+Q_1}{2 B \Delta}\;\delta_- -\frac{P_1}{2B
\epsilon}\;\delta_+)
\end{eqnarray}
\begin{eqnarray}
W_2=-Q_2 \lbrack \epsilon \delta_+(P_2+Q_1)+P_2
\Delta\delta_-\rbrack(\frac{-P_2+Q_1}{2 B \Delta}\;\delta_-
-\frac{P_1}{2B \epsilon}\;\delta_+)
\end{eqnarray}
\begin{eqnarray}
W_3=Q_1 \lbrack \epsilon\delta_+(-P_1+Q_2)-P_1
\Delta\delta_-\rbrack(-\frac{P_1+Q_2}{2 B \Delta}\;\delta_-
-\frac{P_2}{2B \epsilon}\;\delta_+)
\end{eqnarray}
and
\begin{eqnarray}
W_4=Q_1 \lbrack \epsilon\delta_+(P_1+Q_2)+P_1
\Delta\delta_-\rbrack(-\frac{P_1-Q_2}{2 B \Delta}\;\delta_-
-\frac{P_2}{2B \epsilon}\;\delta_+)
\end{eqnarray}
Here $\delta_{\pm}=(\sqrt{\epsilon^2+\Delta^2} \pm \epsilon)/2
C_{\pm}\sqrt{\epsilon^2+\Delta^2}$. Figure 4 displays this
survival probability as well as the transition probabilities
$P_{0L}(t)=\vert \langle 0L \vert 0R \rangle_t \vert^2$ (dashed
line in figure 4a), $P_{1R}(t)=\vert \langle 1R \vert 0R \rangle_t
\vert^2$ and $P_{1L}(t)=\vert \langle 1L \vert 0R \rangle_t
\vert^2$ (dashed line in figure 4b) to the states $\vert 0L
\rangle$, $\vert 1R\rangle$ and $\vert 1L\rangle$ respectively.
Also in this case we underline that the time evolution of all
these transition probabilities is governed by the $4$
characteristic frequencies $\omega_1$, $\omega_2$, $\omega_3$ and
$\omega_4$ previously defined.
\begin{figure}
\includegraphics{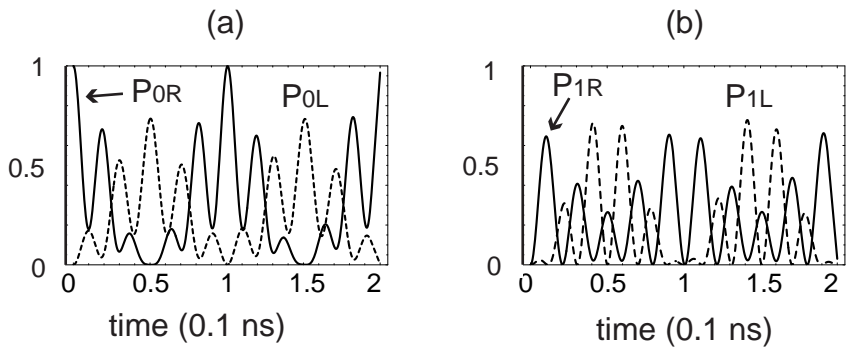}
\caption{\footnotesize \textbf{(a)} Survival probability $P_{0R}$
of the state $\vert 0R \rangle$ and transition probabilities
$P_{0L}$ (dashed line), \textbf{(b)} $P_{1R}$ and $P_{1L}$ (dashed
line) to the states $\vert 0L \rangle$, $\vert 1R \rangle$ and
$\vert 1L \rangle$ respectively, for a system with $\epsilon
\approx 3 \cdot 10^{10} \;rad \cdot s^{-1}, \;B \approx  10^{-34}
J \cdot s $ and $\omega_F \approx 10^{11} rad \cdot s^{-1}  $. }
\end{figure}

This leads to a very rich dynamics of the system, characterized by
the occurrence of entangled states of the total coupled system
obtained by the superposition of states with opposite sense of
circulation of the supercurrent in the loop and a different number
of photons in the field ($0$ or $1$). However, due to the fact
that these characteristic frequencies are not rationally related
to each other, it is impossible to restore exactly  the initial
condition of the system. As we will show in the next section, it
is possible to find an exact correspondence between these
frequencies and then a more regular behavior for the total system,
in the symmetric case and choosing properly the values of the
coupling strength $k$ and of the system parameters.

\subsection{\label{sec:level2} The dynamics of the system in the symmetric case: existence of quantum
 superpositions of clockwise and counterclockwise supercurrent states}
In the previous case the asymmetric SQUID potential configuration
results from the application of a dc control flux $\phi_x$ not
exactly equal to $\phi_0/2$.  In this section we will study the
system when $\phi_x= \phi_0/2$. This means that the two SQUID
potential wells have the same height so that $\epsilon=0$. In such
symmetric conditions hamiltonian \eqref{hasimm} reduces to the
relatively simpler form

\begin{eqnarray}\label{hsimm}
H_R= \left( \begin{array}{c c c c}

0 & 0 & 0 & B \Delta  \\

0 &  \hbar \omega_F & B \Delta & 0\\

0 & B \Delta & \hbar \omega_F & 0 \\

B \Delta & 0 &  0&2 \hbar \omega_F

\end{array}
\right)
\end{eqnarray}
where $B$ must be calculated putting $\epsilon=0$ in eq.
\eqref{B}.

Analyzing the structure of matrix \eqref{hsimm} it is not
difficult to convince oneself that there exist two dynamically
separated subspaces, characterized by the frequencies
$\Omega_1=\frac{B}{\hbar}\; \omega_F$ and
$\Omega_2=\frac{\sqrt{\hbar^2+B^2}}{\hbar} \; \omega_F$,
respectively. The first subspace is generated by $\vert 0+
\rangle$ and $\vert 1- \rangle$ and the representation of $H_R$ on
it is given by the central $2 \times 2$ matrix block. Such a
structure is responsible of the appearance of entanglement in the
time evolution of the combined rf-SQUID-field system. The matrix
elements of $H$ connecting the states $\vert 0- \rangle $ and
$\vert 1+ \rangle$ generating the second subspace reflect the
contribution of Counter-Rotating terms in the truncated version of
H \cite{vietri}.

The eigenstates of matrix \eqref{hsimm} assume the simple form
\begin{eqnarray}\label{u1s}
\vert u_{1s} \rangle=\frac{1}{\sqrt{2}} \lbrack -\vert1- \rangle+
\vert0+ \rangle\rbrack
\end{eqnarray}
\begin{eqnarray}\label{u2s}
\vert u_{2s} \rangle=\frac{1}{\sqrt{2}} \lbrack \vert1- \rangle+
\vert0+ \rangle\rbrack
\end{eqnarray}
\begin{eqnarray}\label{u3s}
\vert u_{3s} \rangle=\frac{1}{\sqrt{n_{3s}}} \lbrack -\frac{(\hbar
+\sqrt{B^2 +\hbar^2 } )}{B} \;\vert 0- \rangle + \vert 1+ \rangle
\rbrack
\end{eqnarray}
\begin{eqnarray}\label{u4s}
\vert u_{4s} \rangle=\frac{1}{\sqrt{n_{4s}}} \lbrack -\frac{(\hbar
-\sqrt{B^2 +\hbar^2 } )}{B} \; \vert 0- \rangle + \vert 1+ \rangle
\rbrack
\end{eqnarray}
with eigenvalues given by
\begin{eqnarray}\label{l1}
\lambda_{1s}= (\hbar-B)\;\omega_F
\end{eqnarray}
\begin{eqnarray}\label{l2}
\lambda_{2s}= (\hbar+B)\; \omega_F
\end{eqnarray}
\begin{eqnarray}\label{l3}
\lambda_{3s}= (\hbar -\sqrt{B^2 +\hbar^2})\;\omega_F
\end{eqnarray}
\begin{eqnarray}\label{l4}
\lambda_{4s}= (\hbar +\sqrt{B^2 +\hbar^2})\;\omega_F.
\end{eqnarray}

Expressions \eqref{u1s}-\eqref{u4s} and \eqref{l1}-\eqref{l4} may
be immediately derived from those relative to the asymmetric case
given in Appendix A in the limit for $\epsilon=0$. The two
eigenstates $\vert u_{1s} \rangle$ and $\vert u_{2s} \rangle$
describe a maximum entangled condition of the rf-SQUID and the
monochromatic field states, induced by the inductive coupling
between them. Rabi oscillations between the degenerate states
$\vert 0+ \rangle$ and $\vert 1- \rangle$ dominate the dynamical
behavior of the system whose time evolution may be written down as
\begin{eqnarray}\label{0+time}
\vert 0+ \rangle_t=\frac{1}{\sqrt{2}} \lbrack \vert u_{1s} \rangle
\exp\;(-i \lambda_{1s}t / \hbar) +\vert u_{2s} \rangle \exp\;(-i
\lambda_{2s}t / \hbar)\rbrack
\end{eqnarray}
if  the initial condition $\vert 0+ \rangle$ is assumed. Also in
this case we are interested in exploring the ability of the system
to periodically come back to the initial state $\vert 0+ \rangle$
as well as to pass through the state $\vert 1-\rangle$. To this
end we plot both the survival probability $P_1(t)=\vert \langle
0+\vert0+ \rangle_t \vert^2=\frac{1}{2}(1+ \cos 2 \Omega_1t)$ of
the initial state (solid line in figure 5a) and the probability
$P_2(t)=\vert \langle 1-\vert0+ \rangle_t \vert^2=\sin^2
\Omega_1t$ to find the system in the state $\vert 1- \rangle$
after a time $t$ (dashed line in figure 5a). Eq. \eqref{0+time}
together with figure 5a provides a clear evidence of the existence
of coherent Rabi oscillations with frequency $\Omega_1$ between
the states $\vert 0+ \rangle$ and $\vert 1- \rangle$ corresponding
to the emission and absorption of a quantum of energy $\hbar
\omega_F$ by the rf-SQUID.

Now let us consider  the $2 \times 2$ subspace in which the
dynamics of the system, with respect to the truncated Hamiltonian
\eqref{hsimm}, is governed only by the Counter Rotating terms with
characteristic frequency $\Omega_2$. Preparing the system in the
state $\vert 0- \rangle$ and considering its time evolution, we
easily get
\begin{eqnarray}\label{0-time}
\vert 0- \rangle_t=-\frac{B^2}{4(\hbar^2+B^2)} \lbrack
\frac{1}{\sqrt{n_{4s}}} \vert u_{3s}\rangle \exp\;(-i \lambda_{3s}
t/ \hbar)-\frac{1}{\sqrt{n_{3s}}} \vert u_{4s}\rangle \exp\;(-i
\lambda_{4s} t/ \hbar)\rbrack.
\end{eqnarray}
Once more we calculate the  survival probability $P_3(t)=\vert
\langle 0- \vert 0- \rangle_t \vert^2$ of the ground state $\vert
0- \rangle$
\begin{eqnarray}
P_3(t)=\frac{B^2}{4(B^2+\hbar^2)}\lbrack  \frac{4
\hbar^2}{B^2}+2(1- \cos 2 \Omega_2t)\rbrack
\end{eqnarray}
and the transition probability $P_4(t)=\vert \langle 1+ \vert 0-
\rangle_t \vert^2$ to the state $\vert 1+ \rangle$
\begin{eqnarray}
P_4(t)=\frac{B^2}{2(B^2+\hbar^2)}\lbrack 1- \cos 2
\Omega_2t\rbrack.
\end{eqnarray}

Analyzing the structure of these two expressions, we deduce that
 the entanglement between the two interacting subsystems leads to
an oscillatory behavior as before but now, as shown also in figure
5b, we cannot get a complete  population inversion between the
ground state $\vert 0- \rangle$ and the higher energy excited
state $\vert 1+ \rangle$. This is due to the fact that, in this
reduced Hilbert space, processes involving the exchange of two
quanta of energy between the two subsystems are unlikely.

\begin{figure}
\includegraphics{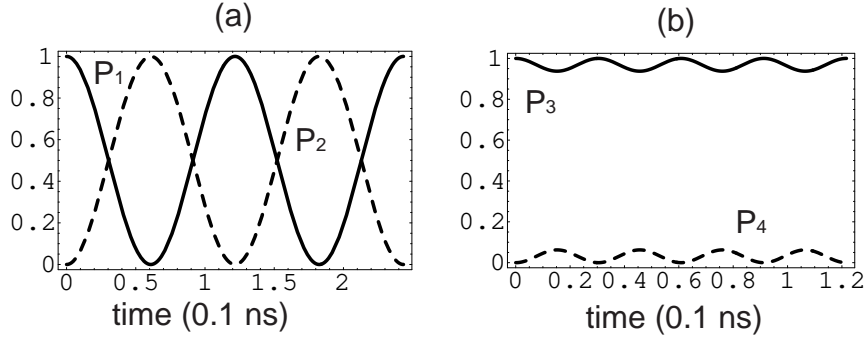}
\caption{\footnotesize  \textbf{(a)} Survival probability $P_1$
 and transition probability $P_2$ (dashed line) at the state
$\vert 1- \rangle$ for a system initially prepared in the state
$\vert 0+ \rangle$.  \textbf{(b)} Survival probability $P_3$ and
transition probability $P_4$ (dashed line) at the state $\vert 1+
\rangle$ for a system initially prepared in the state $\vert 0-
\rangle$. In both cases we set $\epsilon=0 , \;B \approx 2,7 \cdot
10^{-35} J \cdot s $ and $\omega_F \approx 10^{11} rad \cdot
s^{-1}$.}
\end{figure}

Until now we have considered the system initially prepared in a
state belonging to one of the two dynamically independent $2
\times 2$ subspaces. In the following we wish to consider  richer
physical situations involving both the two subspaces at the same
time.

Considering, in fact, as initial condition the state $\vert 0R
\rangle=\vert 0 \rangle \otimes \vert R \rangle$, namely the field
in its vacuum state $\vert 0 \rangle$ and the SQUID with a
right-hand current in the loop, the system evolves in accordance
with the following expression
\begin{eqnarray}\label{0rt}
\vert 0R \rangle_{t}= \alpha(t) \vert 0R \rangle+ \beta(t) \vert
0L \rangle +\gamma(t) \vert 1R \rangle+ \delta(t)\vert 1L \rangle.
\end{eqnarray}
The time dependent parameters appearing in eq. \eqref{0rt} are
linear combinations of trigonometric functions characterized by
the incommensurable frequencies $\Omega_1$ and $\Omega_2$. In this
case the time evolution of the system is rather similar to that
obtained in the asymmetric case. Generally speaking, this fact
makes it impossible for the system to exactly restore its initial
condition. Since the ratio $\Omega_2/ \Omega_1$ may be controlled
by acting upon the parameter $B$ and then, at least, on one of
the physical parameters appearing in its expression given by eq.
\eqref{B}, we may wonder on what the dynamical properties of the
system become in correspondence to special values of $B$. It is
indeed immediate to see that for
\begin{eqnarray} \label{bnuovo}
B=\hbar / \sqrt{n^2-1}\
\end{eqnarray}
we get $\Omega_2=n \; \Omega_1$, where $n>1$ is an arbitrary
integer. Under such a controllable condition the dynamics of the
system is dominated by the occurrence of many interesting
features. The four time dependent parameters appearing in eq.
 \eqref{0rt} assume the form
\begin{eqnarray}\label{alfa}
\alpha(\tau)= \frac{\exp\; (-i \tau \sqrt{n^2-1})}{2} (\cos
\tau+\cos n \tau +i\frac{\sqrt{n^2-1}}{n} \sin n \tau),
\end{eqnarray}
\begin{eqnarray}\label{beta}
\beta(\tau) = \frac{\exp\; (-i \tau \sqrt{n^2-1})}{2} (- \cos
\tau+\cos n \tau+i\frac{\sqrt{n^2-1}}{n} \sin n \tau),
\end{eqnarray}
\begin{eqnarray}\label{gamma}
\gamma(\tau) =-i \frac{\exp\; (-i \tau \sqrt{n^2-1})}{2} (\sin
\tau+\frac{1}{n} \sin n \tau)
\end{eqnarray}
and
\begin{eqnarray}\label{delta}
\delta(\tau) = -i \frac{\exp\; (-i \tau \sqrt{n^2-1})}{2}(\sin
\tau - \frac{1}{n} \sin n \tau),
\end{eqnarray}
where $\tau=\Omega_1t$. Analyzing the time evolution of
these four time dependent probability amplitudes, it is not
difficult to convince oneself that \emph{the system comes right
back to the initial state} $\vert 0R \rangle$ after a time $t_1
 \equiv\frac{2 \pi}{\Omega_1}= \frac{2 \pi
\sqrt{n^2-1}}{\omega_F}$ if $n$ is even and after a time $t_1/2$
if $n$ is odd. For this reason eq. \eqref{bnuovo} expresses the
condition for the occurrence of periodic behavior in the dynamics
of the combined system.

Other interesting manifestations in the dynamic of the system
occur depending on the parity of the ratio between $\Omega_2$ and
$\Omega_1$ determined by the fixed value of $B$ in accordance to
eq. \eqref{bnuovo}.

We find indeed that, if $n$ is even and always starting from the
state $\vert 0R \rangle$, at time $t_1/2$ the combined system
reaches the factorized state wherein the field is still in its
vacuum state $\vert 0 \rangle $ and the current in the loop
reverses its sense of circulation, meaning that the state of the
rf-SQUID becomes $\vert L \rangle$. If $n$ is odd, on the
contrary, the probability $P_{0L}(t)$ of finding the system in the
state $\vert 0L \rangle$ is always less than $1$ (see for example
for $n=3$ the dashed line in figure 6a).

Moreover, for $n$ even, at times $t_1/4$ and $3 t_1/4$ the system
is once more describable in terms of factorized states. For
example for $n=4$ these factorized states may be expressed as
\begin{subequations}\label{psit2 1}
\begin{eqnarray}
\vert \psi_0(t_1/4) \rangle &=& \frac{\exp(-i \pi
\sqrt{15}/2)}{\sqrt{2}}\lbrack \vert 0 \rangle -i \vert 1 \rangle
\rbrack \otimes \vert - \rangle, \\ \vert \psi_0(3 t_1/4) \rangle
&=& \frac{\exp(-i 3 \pi \sqrt{15}/2)}{\sqrt{2}}\lbrack \vert 0
\rangle +i \vert 1 \rangle \rbrack  \otimes \vert - \rangle,
\end{eqnarray}
\end{subequations}
 $\vert - \rangle$ being the ground state of the rf-SQUID.
This coherent evolution is represented in figure 7a where we plot,
for $n=4$, the survival probability $P_{0R}(t)$ (dashed line) of
$\vert 0R \rangle$ and the transition probabilities $P_{0L}(t)$,
$P_{1R}(t)$ (dashed line) and $P_{1L}(t)$ to the states $\vert 0L
\rangle$, $\vert 1R \rangle$ and $\vert 1L \rangle$ respectively.

Analyzing eq. \eqref{0rt} and eqs. \eqref{alfa}-\eqref{delta} with
$n=4$, we find that, starting with the field in the vacuum state,
at the same instants wherein the rf-SQUID get disentangled from
the field the probability of finding the  rf-SQUID in its state
$\vert + \rangle$ exactly vanishes. On the contrary, we may
dynamically obtain this condition preparing the system at $t=0$ in
the state $\vert 1R \rangle$. It is indeed easy to prove that in
this case, always putting $n=4$ in eq. \eqref{bnuovo} the state of
the total system at time $t_1/4$ and $3 t_1/4$ becomes once more
factorizable assuming the following form:
\begin{subequations}\label{psit2 2}
\begin{eqnarray}
\vert \psi_1(t_1/4) \rangle &=& \frac{\exp(-i \pi
\sqrt{15}/2)}{\sqrt{2}}\lbrack \vert 1 \rangle -i \vert 0 \rangle
\rbrack  \otimes \vert + \rangle, \\ \vert \psi_1(3 t_1/4) \rangle
&=& \frac{\exp(-i3 \pi \sqrt{15}/2)}{\sqrt{2}}\lbrack \vert 1
\rangle +i \vert 0 \rangle \rbrack  \otimes \vert + \rangle,
\end{eqnarray}
\end{subequations}
$\vert + \rangle$ being the first excited state
of the SQUID. Also in this case our theory describes coherent
oscillations between the states $\vert 1R \rangle$ and $\vert 1L
\rangle$, that is the inversion of the supercurrent flow in the
loop with period $t_1/2$. This coherent behavior of the system
assuming $\vert 1R \rangle$ as initial condition, is represented
in figure 7b where, in correspondence to $n=4$, we plot the
survival probability $P_{1R}(t)$ (dashed line) of $\vert 1R
\rangle$  and the transition probabilities $P_{1L}(t)$,
$P_{0R}(t)$ (dashed line) and $P_{0L}(t)$ to the states $\vert 1L
\rangle$, $\vert 0R \rangle$ and $\vert0L \rangle$ respectively.
It has to be stressed that, on the contrary, when $n$ is odd, the
previously described oscillations between factorized states and
entangled states of the total matter-radiation system do not
occur. This fact may be fully recognized looking at figure 6,
where we plot $P_{0R}(t)$ $P_{0L}(t)$, $P_{1R}(t)$ and $P_{1L}(t)$
assuming $n=3$ and $P_{0R}(0)=1$. We note that a time instant in
correspondence of which these four probabilities are all equal
doesn't exist.

Remembering that $\vert \pm \rangle = \frac{1}{\sqrt{2}} \lbrack
\vert R \rangle \mp \vert L \rangle \rbrack$ and that the states
$\vert R \rangle$ and $\vert L \rangle$ may be legitimately
considered as macroscopically distinguishable states of the
rf-SQUID, eq. \eqref{psit2 1} (\eqref{psit2 2}) predicts the
generation of a maximally entangled Schr\"{o}dinger cat like state
in the dynamics of a rf-SQUID exposed to a
 single mode quantized electromagnetic field when the combined system is prepared in the state
$\vert 0R \rangle$  $(\vert 1L \rangle)$.

The fact of being able to build quantum superpositions of two states
describing clockwise and counterclockwise supercurrents in the loop,
confirms the role of such nanodevices as simple physical systems
thanks to which it is possible to conceive experiments on fundamental
aspects of the quantum theory.

\begin{figure}
\includegraphics {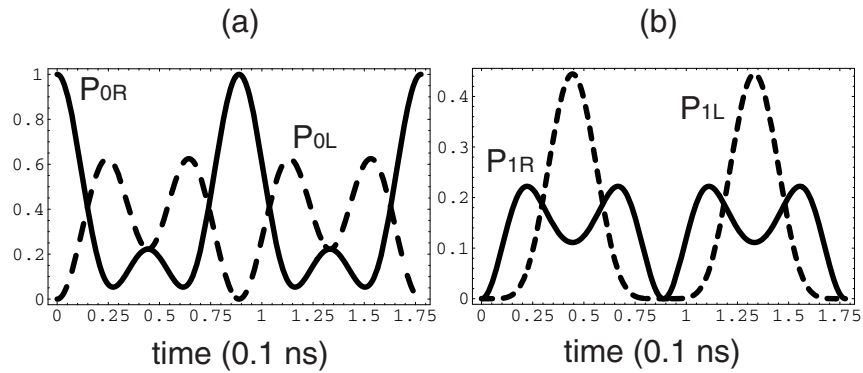}
\caption{\footnotesize \textbf{(a)} Survival probability  of the
state $\vert 0R \rangle$ and transition probabilities to the
states $\vert 0L \rangle$ (dashed line), \textbf{(b)} $\vert 1R
\rangle$ and $\vert 1L \rangle$ (dashed line) for a system
initially prepared in the state $\vert 0R \rangle$. Here, $n=3$,
$\epsilon=0 , \;B= \hbar/ \sqrt{8}$ and $\omega_F \approx 10^{11}
rad \cdot s^{-1}$.}
\end{figure}

\begin{figure}
\includegraphics[width=12 cm]{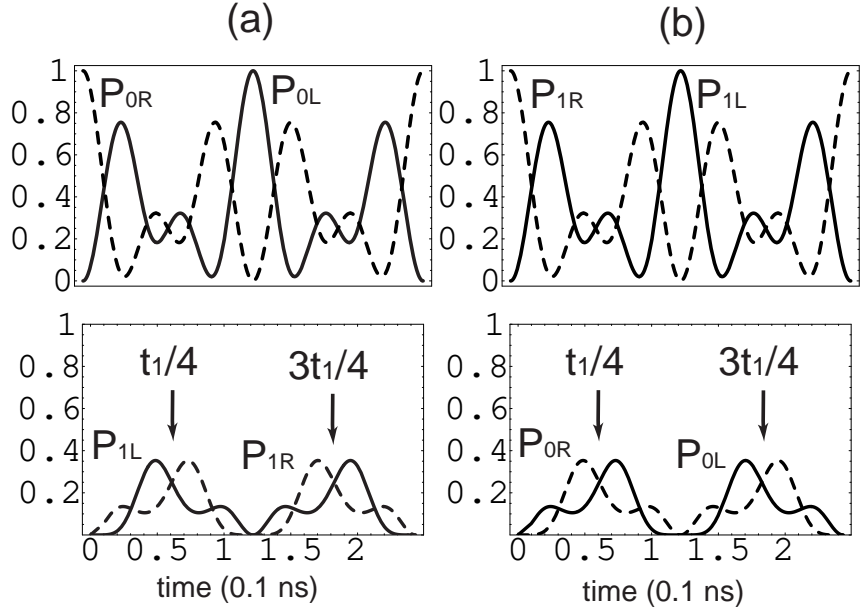}
\caption{\footnotesize \textbf{(a)} Survival probability of the
state $\vert 0R \rangle$ (dashed line) and transition
probabilities to the states $\vert 0L \rangle$, $\vert 1R \rangle$
(dashed line) and $\vert 1L \rangle$ for a system initially
prepared in the state $\vert 0R \rangle$. \textbf{(b)} Survival
probability of the state $\vert 1R \rangle$ (dashed line) and
transition probabilities to the states $\vert 1L \rangle$, $\vert
0R \rangle$ (dashed line) and $\vert 0L \rangle$ for a system
initially prepared in the state $\vert1R \rangle$. Here, $n=4$,
$\epsilon=0 , \;B= \hbar/ \sqrt{15}$ and $\omega_F \approx 10^{11}
rad \cdot s^{-1}$.}
\end{figure}
\section{Discussion and conclusive remarks}

In this paper we have investigated the coupled dynamics of a
rf-SQUID and a single mode quantized electromagnetic field in the
reduced $4 \times 4$ Hilbert space spanned by the low lying energy
states of the uncoupled system. The correspondent Hamiltonian
model includes contributions from both the rotating and
counter-rotating terms and this fact turns out to be at the origin
of a rich dynamical behavior dominated by Rabi oscillations
associated to more than one frequency. By construction, our theory
is based on a Hamiltonian model containing some external
parameters. Since they may be easily varied, we have addressed the
attractive question of the extent at which this circumstance
provides an effective tool to get a reasonable control of some
aspects of the system dynamics. The analysis reported in the paper
considers, fixing appropriate resonance conditions, two different
cases, the asymmetric  and the symmetric ones. In both cases the
dynamical problem is exactly solved in the truncated Hilbert space
finding quasi periodic behaviors of the initial state survival
probability as well as of some physically meaningful transition
probabilities of experimental interest. Such quasi periodic
temporal evolution reflects the existence of quantum coherent
oscillations occurring at incommensurable Rabi frequencies. An
important difference between the two physical situations under
discussion is that in the symmetric case  $H_R$ exhibits two
invariant $2 \times 2$ subspaces whereas in the asymmetric case
the time evolution of any initial condition of the total system
explores the entire $4 \times 4$ Hilbert space. This different
dynamical behavior has direct remarkable consequences. When indeed
the external parameter  $\epsilon$ measuring the height difference
between the two minima of $U(\phi)$ does not vanish,  the system
evolves intrinsically preventing the occurrence of disentangling
in correspondence to any possible choice of  values for the
external parameters. Our theory predicts  completely different and
rich results in the symmetric case stemming from the reducibility
into $2\times 2$ blocks of $H_R$ in the truncated Hilbert space.
We have indeed proved that the external parameters may fixed in
such a way to realize a control on the dynamical replay of the
total system which, for instance, may be forced to exhibit a
periodic evolution accompanied by the occurrence of  an
oscillatory disappearance  of entanglement between the two
subsystems.
A relevant result of this paper, commented at the end of the previous section, is
the generation of quantum superposition of the two macroscopic distinguishable
$\vert L \rangle$ and $\vert R \rangle$ of the rf-SQUID.

These results are new and their value may be further appreciated
considering that the realization of our theoretical scheme is in
the grasp of experimentalists. The several types of SQUID
necessary for the observability  of our predictions are easily
fabricated exploiting the well defined trilayer $Nb/AlO_x/Nb$
technology. Moreover it is possible to prepare and control the
state of the rf-SQUID via flux pulses and rf pulses. Finally we
may readout the macroscopic flux state of the qubit using a
suitable magnetometer (essentially an hysteretic dc-SQUID
detector) whose experimentally measured detection efficiency is of
the order of 98 \% \cite{cosme2}. One of the most crucial problem
related to this kind of device is the unavoidable presence of
decoherence. The quality of coherence for a two-level system can
be qualitatively described in terms of the coherence time
$T_{\varphi}$ of a superposition of its states. It is generally
accepted that for an active decoherence compensation mechanism,
$T_{\varphi}$ must be larger than $10^4t_{op}$, $t_{op}$ being the
duration of an elementary operation of the qubit \cite{preski}. In
our case the Rabi oscillation frequencies $\Omega_1$ and
$\Omega_2$ correspond to characteristic times $t_1 \approx 2.4
10^{-10} sec$ and $t_2 \approx 6 10^{-11} sec$ (for $n=4$).
Moreover, as demonstrated by Cosmelli's group, for a system cooled
at $5 mK$ and effective resistance $R \approx 4 \div 5 M \Omega$,
the decoherence time is approximately of the order of $1 \mu s$
\cite{cosmel}. Thus in our case we can reasonably believe that it
is possible to realize a superconducting device satisfying the
constraint $T_{\varphi}>10^4 t_{op}$. An open problem for the
experimental realization of the physical system is  represented by
the coupling between a qubit and a single-mode of a resonant
cavity. We must take into account the typical dimensions of the
SQUID chip with respect to the size of a high-Q superconducting
cavity. In order to bypass this nontrivial technical problem we
may use an experimental arrangement consisting of a chip placed
inside a cavity made by two open mirrors \cite{walter} or we may
think to integrate the Josephson device and a waveguide in the
same chip \cite{cirill}. Many groups are currently working in this
field and we think (and hope) that these techniques will be of
common use in the next few years. A most immediate solution is
represented by the substitution of the resonant cavity by a LC
resonator or by a large area current-biased Josephson junction.
Several works taking into account this substitution and the fact
that, as discussed in section III, the hamiltonian model for all
these three systems may be expressed in terms of eq.
\eqref{hoscarm}, make it possible to retain that this experiment
may be realized with the currently available technologies.

\begin{acknowledgments}
We wish to acknowledge C. Cosmelli and F. Chiarello for helpful
discussion and F. Intravaia for his technical support. One of the
authors (R.M.) acknowledges financial support from Finanziamento
Progetto Giovani Ricercatori 1999, Comitato 02.
\end{acknowledgments}

\appendix
\section{}

In this appendix we give the analytical expressions for the
eigenstates and eigenvalues of Hamiltonian \eqref{hasimm} in the
asymmetric case. In order to simplify the notation we introduce
the following symbols:
\begin{eqnarray} \label{g}
G=\sqrt{4B^2\epsilon^2+ \hbar^2 \omega_F^2}
\end{eqnarray}
\begin{eqnarray}\label{q1}
Q_1=\sqrt{4B^2 \omega_F^2+2\hbar \omega_F(\hbar \omega_F-G)}
\end{eqnarray}
\begin{eqnarray}\label{q2}
Q_2=\sqrt{4B^2 \omega_F^2+2\hbar \omega_F(\hbar \omega_F+G)}
\end{eqnarray}
\begin{eqnarray}\label{p1}
P_1=\hbar \omega_F+G
\end{eqnarray}
\begin{eqnarray}\label{p2}
P_2=\hbar \omega_F-G
\end{eqnarray}

The eigenvalues of the Hamiltonian \eqref{hasimm} may be written
down as follows:
\begin{eqnarray}\label{la1}
\lambda_1=\hbar \omega_F- \frac{Q_1}{2}
\end{eqnarray}
\begin{eqnarray}\label{la2}
\lambda_2=\hbar \omega_F+ \frac{Q_1}{2}
\end{eqnarray}
\begin{eqnarray}\label{la3}
\lambda_3=\hbar \omega_F- \frac{Q_2}{2}
\end{eqnarray}
\begin{eqnarray}\label{la4}
\lambda_4=\hbar \omega_F+ \frac{Q_2}{2}.
\end{eqnarray}

The eigenstates $\vert u_1 \rangle$, $\vert u_2 \rangle$, $\vert
u_3 \rangle$ and $\vert u_4 \rangle$ relative to the eigenvalues
$\lambda_1$, $\lambda_2$, $\lambda_3$ and $\lambda_4$ respectively
assume the following form:
\begin{eqnarray}\label{u1}
\vert u_1 \rangle= \frac{1}{\sqrt{n_1}} \lbrace -
\frac{Q_1+P_2}{2B \Delta}\;\vert 0- \rangle +
\frac{P_1Q_1-4B^2\epsilon^2}{4B^2 \Delta \epsilon } \; \vert1-
\rangle- \frac{P_1}{2B \epsilon}\;\vert0+ \rangle+\;\vert1+
\rangle \rbrace
\end{eqnarray}

\begin{eqnarray}\label{u2}
\vert u_2 \rangle= \frac{1}{\sqrt{n_2}} \lbrace \frac{Q_1-P_2}{2B
\Delta}\;\vert 0- \rangle - \frac{P_1Q_1+4B^2\epsilon^2}{4B^2
\Delta \epsilon } \; \vert1- \rangle- \frac{P_1}{2B
\epsilon}\;\vert0+ \rangle+\;\vert1+ \rangle \rbrace
\end{eqnarray}

\begin{eqnarray}\label{u3}
\vert u_3 \rangle= \frac{1}{\sqrt{n_3}} \lbrace -
\frac{Q_2+P_1}{2B \Delta}\;\vert 0- \rangle +
\frac{P_2Q_2-4B^2\epsilon^2}{4B^2 \Delta \epsilon } \; \vert1-
\rangle- \frac{P_2}{2B \epsilon}\;\vert0+ \rangle+\;\vert1+
\rangle \rbrace
\end{eqnarray}

\begin{eqnarray}\label{u4}
\vert u_4 \rangle= \frac{1}{\sqrt{n_4}} \lbrace \frac{Q_2-P_1}{2B
\Delta}\;\vert 0- \rangle - \frac{P_2Q_2+4B^2\epsilon^2}{4B^2
\Delta \epsilon } \; \vert1- \rangle- \frac{P_2}{2B
\epsilon}\;\vert0+ \rangle+\;\vert1+ \rangle \rbrace
\end{eqnarray}
where $1 / \sqrt{n_i}$, with $i \equiv 1,2,3,4$, are the
normalizing factors satisfying $\langle u_i\vert u_j
\rangle=\delta_{ij}$. It is useful to expand the states $ \vert 0-
\rangle$, $\vert 1- \rangle$, $\vert 0+ \rangle$ and $\vert 1+
\rangle$ in terms of $\vert u_1 \rangle$, $\vert u_2 \rangle$,
$\vert u_3 \rangle$ and $\vert u_4 \rangle$. Inverting eqs.
\eqref{u1}-\eqref{u4} we get:
\begin{eqnarray}\label{0-a}
\vert 0- \rangle=\frac{B \Delta}{(P_1-P_2)Q_1Q_2} \lbrack
\sqrt{n_1}P_2Q_2 \vert u_1 \rangle - \sqrt{n_2} P_2Q_2 \vert u_2
\rangle\rbrack -\sqrt{n_3} P_1Q_1 \vert u_3 \rangle+ \sqrt{n_4}
P_1Q_1 \vert u_4 \rangle \rbrack
\end{eqnarray}
\begin{eqnarray}\label{1-a}
\vert 1- \rangle=\frac{2B^2 \Delta \epsilon}{(P_1-P_2)Q_1Q_2}
\lbrack \sqrt{n_1}Q_2 \vert u_1 \rangle - \sqrt{n_2} Q_2 \vert u_2
\rangle\rbrack -\sqrt{n_3} Q_1 \vert u_3 \rangle+ \sqrt{n_4} Q_1
\vert u_4 \rangle \rbrack
\end{eqnarray}
\begin{multline}\label{0+a}
\vert 0+ \rangle=\;\frac{B \epsilon}{(P_1-P_2)Q_1Q_2}\;\lbrack
\sqrt{n_1}(P_2-Q_1)Q_2 \;\vert u_1 \rangle
-\sqrt{n_2}(P_2+Q_1)Q_2\;\vert u_2 \rangle
\\-\sqrt{n_3}(P_1-Q_2)Q_1 \;\vert u_3 \rangle +
\sqrt{n_4}(P_1+Q_2)Q_1 \;\vert u_4\rangle \rbrack
\end{multline}
\begin{multline}\label{1+a}
\vert 1+ \rangle=\;\frac{1}{2(P_1-P_2)Q_1Q_2}\;\lbrack \sqrt{n_1}
\; Q_2(P_1P_2+P_2^2-P_2Q_1+4B^2 \epsilon^2)\;\vert u_1 \rangle +
\\
-\sqrt{n_2} \;Q_2 (P_1P_2+P_2^2+P_2Q_1+4B^2 \epsilon^2)\;\vert u_2
\rangle
\\-\sqrt{n_3}\;Q_1 (P_1^2+P_1P_2-P_1Q_2+4B^2 \epsilon^2)   \;\vert u_3 \rangle +
\sqrt{n_4} \; Q_1(P_1^2+P_1P_2+P_1Q_2+4B^2 \epsilon^2) \;\vert
u_4\rangle \rbrack.
\end{multline}

\section{}

In this section we give the analytical expressions for the
transition probabilities $P_2(t)$, $P_3(t)$ and $P_4(t)$ to the
states $\vert 1- \rangle$, $\vert 0- \rangle$ and $\vert 1+
\rangle$ for a system in the asymmetric configuration and prepared
at $t=0$ in the state $\vert 0+ \rangle$.

Exploiting eqs. \eqref{0+asimmt}, \eqref{0-a}, \eqref{1-a} and
\eqref{1+a}, it is immediate to written down these transition
probabilities as:
\begin{multline}\label{P1-asimm}
P_2(t)=\vert \langle1-\vert0+ \rangle_t \vert^2=S^2 \lbrace
\sum^4_{j=1}T_j^2+2T_1T_2 \; \cos\frac{Q_1}{\hbar}t+2T_3T_4 \;
\cos \frac{Q_2}{\hbar}t+\\+2 \lbrack T_1T_3+T_2T_4\rbrack \;\cos
\frac{(Q_1-Q_2)}{2 \hbar}t+ 2 \lbrack T_2T_3+T_1T_4\rbrack \; \cos
\frac{(Q_1+Q_2)}{2\hbar}t\rbrace
\end{multline}
where
\begin{eqnarray}\label{T1}
T_1=Q_2 \frac{P_2-Q_1}{4B^2 \epsilon \Delta}(-4B^2
\epsilon^2+Q_1P_1)
\end{eqnarray}
\begin{eqnarray}\label{T2}
T_2=Q_2 \frac{P_2+Q_1}{4B^2 \epsilon \Delta}(4B^2
\epsilon^2+Q_1P_1)
\end{eqnarray}
\begin{eqnarray}\label{T3}
T_3=Q_1 \frac{-P_1+Q_2}{4B^2 \epsilon \Delta}(-4B^2
\epsilon^2+Q_2P_2)
\end{eqnarray}
and
\begin{eqnarray}\label{T4}
T_4=Q_1 \frac{P_1+Q_2}{4B^2 \epsilon \Delta}(-4B^2
\epsilon^2-Q_2P_2).
\end{eqnarray}
Also,
\begin{multline}\label{P0-asimm}
P_3(t)=\vert \langle0-\vert0+ \rangle_t \vert^2=S^2 \lbrace
\sum^4_{j=1}Z_j^2+2Z_1Z_2 \; \cos\frac{Q_1}{\hbar}t+2Z_3Z_4 \;
\cos \frac{Q_2}{\hbar}t+\\+2 \lbrack Z_1Z_3+Z_2Z_4\rbrack \;\cos
\frac{(Q_1-Q_2)}{2 \hbar}t+ 2 \lbrack Z_2Z_3+Z_1Z_4\rbrack \; \cos
\frac{(Q_1+Q_2)}{2\hbar}t\rbrace
\end{multline}
where
\begin{eqnarray}\label{Z1}
Z_1=Q_2 \frac{P_2-Q_1}{2B  \Delta}(-P_2-Q_1)
\end{eqnarray}
\begin{eqnarray}\label{Z2}
Z_2=-Q_2 \frac{P_2+Q_1}{2B \Delta}(-P_2+Q_1)
\end{eqnarray}
\begin{eqnarray}\label{Z3}
Z_3=Q_1 \frac{-P_1+Q_2}{2B  \Delta}(-P_1-Q_2)
\end{eqnarray}
\begin{eqnarray}\label{Z4}
Z_4=Q_1 \frac{P_1+Q_2}{2B  \Delta}(-P_1+Q_2)
\end{eqnarray}
and
\begin{multline}\label{P1+asimm}
P_4(t)=\vert \langle1+\vert0+ \rangle_t \vert^2=S^2 \lbrace
\sum^4_{j=1}X_j^2+2X_1X_2 \; \cos\frac{Q_1}{\hbar}t+2X_3X_4 \;
\cos \frac{Q_2}{\hbar}t+\\+2 \lbrack X_1X_3+X_2X_4\rbrack \;\cos
\frac{(Q_1-Q_2)}{2 \hbar}t+ 2 \lbrack X_2X_3+X_1X_4\rbrack \; \cos
\frac{(Q_1+Q_2)}{2\hbar}t\rbrace
\end{multline}
where
\begin{eqnarray}\label{X1}
X_1=Q_2 (P_2-Q_1)
\end{eqnarray}
\begin{eqnarray}\label{X2}
X_2=-Q_2 (P_2+Q_1)
\end{eqnarray}
\begin{eqnarray}\label{X3}
X_3=Q_1 (-P_1+Q_2)
\end{eqnarray}
\begin{eqnarray}\label{X4}
X_4=Q_1 (P_1+Q_2).
\end{eqnarray}

\newpage
\bibliography{apssamp}% Produces the bibliography via BibTeX.

\end{document}